# Noise Analysis and Detection Based on RF Energy Duration in wireless LAN

Dr.R.Seshadri [1] and Prof.N..Penchalaiah [2]

[1]Prof & Director of university computer center, S.V.University, Tirupati, India
`ravalaseshadri@gmail.com`
[2]Department of Computer Science Engineering, ASCET, Gudur,Andhra pradesh,India
`pench_n@yahoo.com`

*Abstract*

*Noise is the major problem while working with wireless LAN. In this paper we analyze the noise by using active receiving antenna and also propose the detection mechanism based on RF energy duration. The standard back off mechanism of 802.11 wireless LAN (WLAN) increases the contention window when a transmission failure occurs in order to alleviate contentions in a WLAN. In addition, many proposed schemes for 802.11 WLAN behave adaptively to transmission failures. Transmission failures in WLANs occur mostly by two causes: collision and channel noise. However, in 802.11 WLAN, a station cannot know the cause of a transmission failure, thus the adaptive schemes assume the ideal situation in which all transmission failures occur by only one of two causes. For this reason, they may behave erroneously in a real world where transmission failures occur by both causes. In this paper, we propose a novel scheme to detect collision, which utilizes transmission time information and RF energy duration on the channel. By detecting collisions, a station can differentiate the causes of transmission failures and the adaptive schemes can operate correctly by using the detection information.*

## 1. INTRODUCTION

With science and technology developed quickly. The active antenna has been getting more and more important. By the word "active antenna ", we mean that the conducting wire of an antenna is combined with an active device to form a non-separated unit. Its advantage is small size wide frequency band, and so on. The reason of its miniaturization is that the gain decrement due to the reduction in the size of the antenna wire is compensated by the gain of the active network. As soon as the impedance of the antenna wire is level-matched with the one of the active network, the antenna impedance is extremely non-sensitive to the frequency, which leads to a wire frequency band. Increasing active network bring about the extra noise, which limits decreasing the size of the antenna and extending its frequency band. Therefore, the noise suppression becomes the bottleneck in technique. Thus, measuring the noise figure is very important. So far, as we know, the norm to measure noise figure of an active antenna has not been suggested. For some active antennas, the noise figure of the active network is only given. In this paper, the noise of the active antenna and its noise figure are analyzed. A method to measure the noise figure of active antennas is developed which benefits the research and making of the active antennas.

As the demand for high-speed wireless communication increases, wireless local area network (WLAN) is emerging as a promising solution. Especially, IEEE 802.11 [1] is the most popular WLAN technology, which supports high data rates up to 54 Mbps in the ISM bands such as 2.4 GHz and 5 GHz. The IEEE 802.11 standard specifies two Medium Access Control (MAC) schemes: a mandatory Distributed Coordination Function (DCF) and an optional Point Coordination Function (PCF). Currently, most of WLAN devices implement DCF only due to its simplicity and efficient best-effort service provisioning. DCF is based on carrier sense multiple access with collision avoidance (CSMA/CA), in which a station transmits its frame only if the medium is determined to be empty, i.e., no other stations transmit. The collision avoidance mechanism utilizes the random back off prior to each frame transmission attempt.

Transmission failures in the IEEE 802.11 WLAN occur mostly by two causes: collision and channel noise. While the random back off can reduce the collision probability, it cannot completely eliminate the collisions since two or more stations can finish their back off procedures simultaneously. A transmission attempt can also fail without a collision since the wireless channel I error-prone due to path loss, interference, etc.. We refer to such transmission failures as a channel error to differentiate it from a collision. Although transmission failures occur by two causes, wireless station cannot realize the causes since, unlike wiredLANs, the collision detection mechanism cannot be implemented in WLANs. Instead, the stations know only whether the transmitted frame has been received successfully by acknowledgement (ACK). For this reason, many operation principles of 802.11 WLAN and other proposed schemes for 802.11 which behave adaptively to transmission failures assume the ideal situation in which all transmission failures occur by only one cause: collision or channel error. Therefore, the adaptive schemes may behave erroneously and in an undesired manner in the real world where transmission failures result from both causes. Some of them (e.g. [2][3]) assume that all transmission failures are due to collisions. As an example, in the back off mechanism of 802.11 MAC, a station doubles its contention window (CW) value to reduce collision when a transmission failure occurs. This behavior is based on the assumption that transmission failures result from collisions. However, transmission failures can occur also by channel errors. Therefore, for channel errors, the back off mechanism wastes the bandwidth and increases transmission delay unnecessarily. Others (e.g. [4] [5]) assume that all transmission failures are caused by channel errors. One of the examples is the Automatic Rate Fallback (ARF) algorithm [4] used in WaveLAN-I1 products from Lucent, which is a simple link adaptation algorithm. In ARF, a station lowers its transmission bit rate when the certain number of transmission failures occurs consecutively assuming that transmission failures are originated from channel errors. Therefore, if collisions occur frequently due to many contending stations, the stations even in a good channel quality may transmit data frames in an unnecessarily low rate. In order that the adaptive schemes can behave correctly and the performance of 802.11 MAC can be optimized further, a station should be able to differentiate collisions and channel errors. Collision detection is one of the possible approaches to differentiate collisions and channel errors. There has been a notable attempt to detect collisions in sensor networks [6].

This scheme considers two types of collisions in the presence of capture: stronger-first and stronger-last in which the packet with the stronger power comes first and last, respectively. In stronger-first collisions, a receiving node can detect a collision by finding a new extra termination symbol, while, in stronger-last collisions, a receiving nod can detect a collision by finding a new preamble, during the reception of another packet. However, for successful detection, the transmissions which result in a collision should have enough differences I transmission start time and receiving power. Therefore, this scheme can be only applied to restricted collision situations to enable collision detection in various situations of 802.11 WLAN; we propose a novel collision detection scheme in this paper.

## 2. THE NOISE ANALYSIS OF ACTIVE RECEIVING ANTENNA

The antenna system of a receiver is degraded by various noises in its practical applications. The key parameter of a receiver system is its noise figure that is measurement of the degradation in SNR. For an active receiving antenna, the receiver system can be shown as Fig. 1. Its output SNR is

$$\frac{S}{N} = \frac{S}{N_A + N_a + N_r} = \frac{S/N_A}{1 + N_a/N_A + N_r/N_A} \quad (1)$$

Where $N_A$ is the noise power propagated by the ambient noise into the output of the receiver; $N_a$ is the noise power resulted by the inherent noise of the receiver; $N_r$ is the output noise power of the receiver

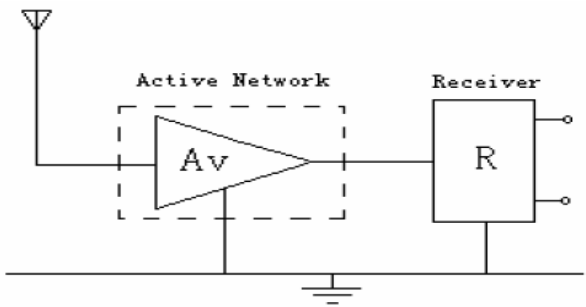

According to the definition of the noise figure, it is know that

$$F = \frac{(S/N)_{in}}{(S/N)_{out}} \quad (2)$$

As compared to Eq. (1), the noise figure of the receiver system becomes

$$F_S = 1 + \frac{N_a}{N_A} + \frac{N_r}{N_A} \quad (3)$$

When $N_A = KT_A BG_r G_a$; $N_a = KT_a BG_r G_a$; $N_r = KT_r BG$; $T_A = T_O(F_A - 1)$, $T_a = T_O(F_a - 1)$, $T_r = T_O(F_r - 1)$, are substituted into Eq. (3), we know

$$F_S = 1 + \frac{F_a - 1}{F_A - 1} + \frac{F_r - 1}{(F_A - 1)G_a} \quad (4)$$

Where $F_S$ is the noise figure of the receiver system when source temperature is the ambient noise temperature $T_A$, $F_A$ is the ambient noise figure, $F_a$ is the one of the active antenna when the source temperature is room-temperature 290K, $F_r$ is the one of receiver when the noise temperature is also room-temperature 290K, and $G_a$ is power gain of the active antennas. $F_A$, $F_a$ and $F_r$ are all called as the normal noise figure. Eq. 4 represents the relationship of the noise figure of receiver system to the ambient noise figure, the noise figure of the active antenna and the noise figure of the receiver from which it is seen that the ambient noise figure plays an important rule in the noise figure of the receiver system. The ambient noise figure includes the cosmic noise figure generalized by the radio-sources in the cosmic space, the radiations from sun spots, the Milky Way, and the cosmic rays, the atmosphere noise mainly resulted from the lightning radiation and the thunder-gust, and artificial noise, e.g., the mechanical and electrical interference, and ambient thermal radiation etc. It is well know that the cosmic noise and atmosphere noise are violently changed with the season, and the place, and also with the frequency. The noise somewhere and its functional relation to the frequency are mainly referred to Report 332 of CCIR. The distribution of the ambient noise and its relation to the frequency in the discussed frequency range are shown in the following figure.

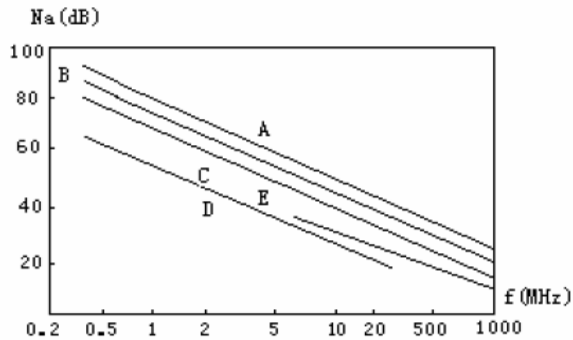

The design of an active antenna is traditionally based on the way in which the characteristic curve of D to E that is inherently a statistically average one is the start of the sensitivity design. Of course, when it is used in a specific range, the specific curve should be applied as the design basis. In short wave band, the ambient noise is more severe. In the case, we request the system noise figure is mainly dependent on the ambient noise figure, that is, $F_s \leq 2$ We know that

$$F_a \leq F_A - \frac{F_r - 1}{G_a} \qquad (5)$$

This is just the request for the noise figure of active receiving antennas. From Eq. (5), it is known that the $G_a$ should be as large as possible. But, considering the nonlinear distortion, $G_a$ cannot be increased unlimited. The working characteristic curve requested by the noise figure of an active receiving antenna in short wave band is shown in Fig.3. The curve is referred to the minimum of the ambient noise level at 30 MHz, and provided that the noises figure of the receiver is 10 dB. Generally, the noise figure of a receiver is not the same, when the used receiver is not the same. Normally, $F_r$ is in the range of 9 to 16 dB. From Fig.3 it is also seen that when $F_r$ is constant, $G_a$ has little effect on $F_a$ that is mainly due to that in short wave band $F_a$ is much larger than $F_r$ so that the effect of the noise figure of the receiver on the noise figure of the receiving system is negligible in general. So, when the noise figure of the active antenna system is substituted into the noise figure of the receiving system, the latter is changed to

$$F_S \approx 1 + \frac{F_a - 1}{F_A - 1} = F_{SA} \qquad (6)$$

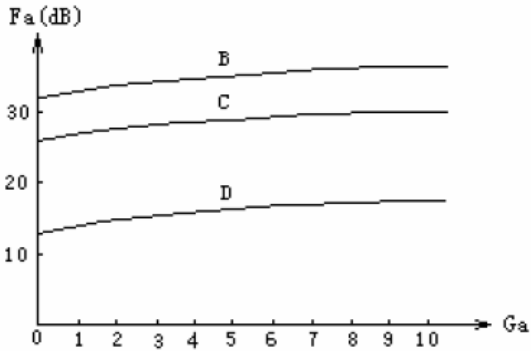

Practically, the noise figure of the active antenna receiving system is just the noise figure of the active antennas that really maps the relation of the output SNR of the active antennas to its input SNR. The word "system" is increased to the term to be discriminated from the traditionally termed noise figure of the active antennas when the source temperature is the room temperature. Eq. (6) gives us the formula evaluating the noise performance of an active antenna.

In an active antenna, increasing an active network in which the noise is basically resulted from its active devices brings about the extra noise. In order to reduce the noise, everything possible must be done to noise-match the conducting wire of the antennas to the active network But, it is impossible to noise-match the monopole antenna work in a wide frequency band even through the exciting network is arbitrarily sophisticated. In order to realize wide frequency band, the impedance-level-match is wildly applied between the antenna wire and the active network, which is obtained from impedance-off-matching and noise-off-matching. The more the off-matching is, the wider the frequency is for a monopole electric-short active antenna, an FET is used as the input stage at most time in active network. In this case, the inherent noise of the FET is less. But, in the electric-short antenna wire that is a source with low resistance and high capacitive reactance, the effective power gain is generally less than 1. Thus, the noise of the second stage must be carefully considered. Everything must be done to noise-match as possible. In order for the noise of the active network to be basically dependent on the first stage, the device with low noise high cutoff frequency and high gain should be selected. Furthermore, the connection mode should also be correctly selected. As far as the noise is only concerned, the common source connection comes into favor,. But, consider the nonlinear distortion and the noise-match of the $2^{nd}$ stage; the common drain connection is more optimal. In a word, a well-done design requests that the noise of the active network should be basically determined by stage 1.

## 3. COLLISION DETECTION BASED ON RF ENERGY TIME

In this section, we describe the collision detection scheme based on RF energy time, named CD-ET. Before description, we define two time values to specify an event in a time axis: ST and DT.

. ST (Start Time): the time point when the event starts to occur.
. DT (Duration Time): the duration for which the event lasts.

Using this two tuple, we can specify transmission attempts of stations and RF energies on the channel in a time axis. We refer to the two tuple of (ST, DT) of a transmission attempt as TT (Transmission Time information). Especially for a RF energy, we refer to its ST and DT as EST and EDT, respectively, and the two tuple of (EST, EDT) as ET (Energy Time information). First, we briefly explain the key idea of CD-ET. When stations start to transmit their frames simultaneously in an infrastructure BSS, a collision occurs and the AP of the BSS sees a merged RF energy caused by the collided frames. If the DTs of the collided frames are different from each other, the EDT of the merged RF energy is longer than the DT(s) of the original frame(s) except the longest frame(s). Therefore, if the sending stations can know the EDT of the RF energy, they can detect the collision by comparing the EDT with the DT of their own transmission attempt. When the STs of the collided frames are different from each other, the stations can detect the collision in the same manner. To enable CD-ET in 802.11 WLAN, we assume that the V/LAN Network Interface Card (NIC) can measure the duration of a RF energy on the channel and report it to the MAC layer.

The measurement of RF energies on the channel is being used by some 802.11 V/LAN NICs to check whether the channel is busy or to check interference in the channel [7]. Since the ambient channel noise changes depending on the environment, the outlier detection algorithm [8] can be used for the accurate measurement of the valid RF energies which are caused by frame transmissions. We also assume that each station stores the TTs of its transmission attempts in a queue TQ' based on its local clock so that it can compare an EDT with DTs of its previous transmission attempts. Fig. 1 shows the detailed operation of CD-ET when a collision occurs by the simultaneous transmission starts of the two stations A and B in an infrastructure BSS. In this figure, we assume that the DT of the collided frame from the station B is longer than that from the station A. With the collision, the AP of the BSS can detect a RF energy on the channel without the successful reception. Then, the AP broadcasts an ET frame, which is a management frame newly define in CDET, in SIFS (Short Inter-Frame Space) after the end of the energy. The ET frame has the same frame format as the beacon frame except that it contains an ET. Before the transmission of the ET frame, the EST is recalculated to EST' based on

$$EST' = EDT + SIFS + T[ET]$$

Where T[ET] is the transmission time of the ET frame. From the equation above, EST' is the time from the start of the RF energy to the end of the transmission of the ET frame. Then, the stations A and B can obtain the start time of the RF energy based on their own local clock by subtracting the received EST' from their local clock time. As shown in Fig.1, the station A can obtain ESTA, which is the start time of the RF energy based on A's local clock, by subtracting the received EST' from its local clock t' (tl is A's local clock time when it receives the ET frame). In this manner, stations in the BSS can share EST without global time synchronization. When the stations receive the ET from the AP, they check the ET against TTs in their TQ. If

62

the EDT is larger than the DT of the transmission attempt which overlaps the ET, the station (A) detects the collision because it comes to know that there was another transmission attempt(s). Here, ET= (EST, EDT) and TT= (ST, DT) are defined to overlap with each other when they satisfy the below condition:

$$(EST < ST < EST+EDT) \text{ OR } (ST < EST < ST+EDT),$$

Which means the RF energy is (possibly partly) caused by the corresponding transmission attempt. This is the end of the first detection phase. The time required for the first detection phase is bounded to the time below:

$$(\max \{ST_j + DT_i\} - \min\{ST_J\}) + SIFS + T[ET]$$

Where $ST_i$ and $DT_i$ is the ST and DT of the i-th overlapping transmission, respectively. At the end of the first detection phase, the station B cannot know the collision yet since the EDT has the same value as the DT of the collided frame from the station B. To notify All the queues considered in this paper are First-In First-Out (FIFO) queues.

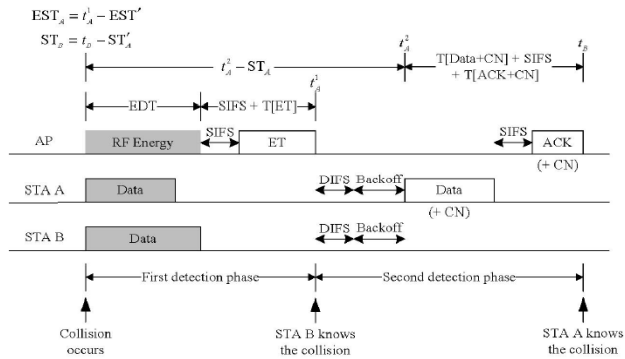

The station B of the collision, the station A piggybacks the collision notification (CN) on its next data frame (Data+CN frame), which contains the STA and DT of its collided frame. Before transmission, the STA of CN should be also recalculated to ST' in the same manner for the ET frame as below:

$$ST' = t - STA + T[Data + CN] + SIFS + T[ACK + CN]$$

Where t2 is A's local clock time when A starts to send the Data+CN frame, and T[Data+CN] and T[ACK+CN] are the transmission times of the corresponding frames. Then, the AP sends the ACK frame with that CN (ACK+CN frame) for the station B since the station B can be hidden from the CN sending station. When B overhears the ACK+CN frame, B calculate $ST_B$, which is the start time of the A's frame based on B's local clock, by subtracting the received ST' from its local clock $t_B$ ($t_B$ is B's local clock time when B receives the ACK+CN frame). If the TT of the received CN and one of the transmission attempts in TQ overlap with each other, the station B considers it a collision since those two transmission attempts interfere with each other. The definition of the overlap of two TTs is the same as that of ET and TT. This is the end of the second detection phase and, at this time; all the collided stations come to know the collision. A collision between more than two stations can be detected in the same manner. CN can be also transmitted by a separate frame, but piggybacking is more efficient because inter frame space and back off duration is a big overhead in 802.11 WLAN. In the real implementation, CN is buffered in the queue CQ since other collisions can occur before the detection finishes. A station piggybacks CN on its data frame whenever its CQ is not empty.

When a station receives CN, it removes the overlapping entry in CQ because that information is already shared by all the stations in the BSS. On the other hand, if the DTs of all the collided frames are equal, CD-ET cannot detect the collision (all their STs are assumed to be equal). This is especially true when all the stations operate in RTS/CTS mode. For this reason, we propose to use Random Bit Padding (RBP) in which each station pads random number of bits at the end of a data frame (RTS frame in RTS/CTS mode) before transmission. By doing this, the length of data frames has randomness. The unit of padding bits is Wz Unit and Wu, it/r should be larger than the minimum time granularity tg which stations can measure (e.g. lps). r is the transmission bit rate. Assuming that tg is constant, Wu unit can be configured adaptively to r to reduce the overhead (e.g. Wz Unit = [tg r]). The number of Wu unit to pad is uniformly chosen in the range of [0, RW-1], where RW is the RBP window. As RW increases, it is more probable that transmitted frames have different duration times. Therefore, the detection capability is determined by the size of RW.Meanwhile, if a Data+CN frame collides with a frame which has the longer DT, the station of the Data+CN frame will have a new CN in its CQ. Then, the new CN should wait for the current Data+CN frame to be transmitted successfully before its transmission, which results in the increase of the detection delay. To alleviate this problem, a station chooses the number of Wunit to pad in the range of [RW, 2RW - 1] for Data+CN frames2. Then, it is more probable that Data+CN frames have the longer DT than normal data frames. Therefore, the station of a normal data frame may have a new CN in a collision assuming that many data frames have the same transmission time. In the presence of capture, the AP does not broadcast the ET frame since it receives one of the collided frames successfully. Therefore, the stations of the non-captured frames cannot detect the collision. For this reason, the AP manages the queue RQ which stores TTs of the successfully received frames. If a station does not receive either ACK and ET frame after it sends a data frame, it piggybacks the failure notification (FN) on the next data frame (Data+FN frame) to inform the AP of its transmission failure. The FN contains the TT of the failed transmission. Then, the AP checks the FN against TTs in RQ, and sends an ACK+CN frame directly if it finds any TT which overlaps with the FN.

## 4. CONCLUSIONS

In this paper, we analyse the noise problem by using active receiving antenna and also we proposed the novel collision detection scheme, called CD-ET, for IEEE 802.11 WLAN for this noise problem. Our scheme can be used with the wide range of adaptive schemes so that the adaptive schemes behave correctly against transmission failures.

International Journal of Distributed and Parallel Systems (IJDPS) Vol.2, No.5, September 2011

**Authors**

**Prof.N.Penchalaiah** Research Scholar in SV University, Tirupati and Working as Professor in CSE Dept,ASCET,Gudur. He was completed his M.Tech in Sathyabama University in 2006. He has 11 years of teaching experience. He guided PG & UG Projects. He published 2 National Conferences and 6 Inter National Journals.

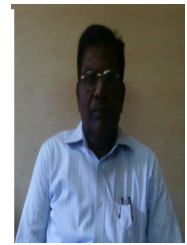

**Dr.R.Seshadri** Working as Professor & Director, University Computer Centre, Sri Venkateswara University, Tirupati. He was completed his PhD in S.V.University in 1998 in the field of " Simulation Modeling & Compression of E.C.G. Data Signals (Data compression Techniques) Electronics & Communication Engg.". He has richest of knowledge in Research field, he is guiding 10 Ph.D in Fulltime as well as Part time. He has vast experience in teaching of 26 years. He published 10 national and

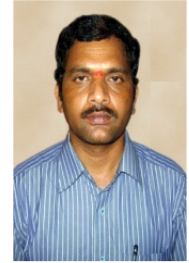